\documentstyle[aps,amssymb]{revtex}  
\def\sech\mbox{{sech\,}}  
\def\beq#1{\begin{eqnarray}\label{#1}}  
\def\eeq{\end{eqnarray}}  

\begin{document}  

\begin{center}  
\null  
%\vskip -1.2in  
  
\begin{flushright}
CU-TP-984
\end{flushright}
\vspace{1cm}  

{\LARGE \bf A Note on the Gauge Group  
of the  \\  
\vskip .2cm  
Electroweak Interactions}  
  
\bigskip  
 V.~Rizov$^a$ and  I.~Vitev$^{a,b}$  
\bigskip

{\sl $^a$ Department of Theoretical Physics, Sofia University St.~Kliment Ohridski, \\  
  
boul. J. Bourchier 5, 1126 Sofia, Bulgaria}\\  
  
{\sl  $^b$ Department of Physics, Columbia University, 538 W 120-th  
Street, \\ New York, NY 10027, USA}

\medskip  
  
\end{center}

\begin{abstract}  
  
We propose a three-fold covering of the group $\mbox{U} (2)$ as a  
gauge group for the electroweak interactions for the purpose of  
describing fields with integer and fractional electric charges with  
respect to the residual electromagnetic gauge group after a  
spontaneous breaking  of the gauge symmetry.  
In a more general scheme we construct a  
three-fold covering of $\mbox{U} (n)$ and consider for the case  
$n=2$  several representations which are used in the construction of  
a model of the electroweak interactions in a subsequent paper.  
\\ 
\vskip .1cm 
\noindent PACS number(s): 02.20.-a; 11.15.-q; 12.15.-y  
\end{abstract}  
%\newpage  

\section{Introduction}  
In the Weinberg-Salam (WS) model \cite{We,Sa} the transformation  
laws under the group $\mbox{U}(1)$ of the weak hypercharge  
$\mbox {Y}$ are different for the quark and lepton fields. Among the  
irreducible representations of $\mbox{U}(1)$, namely  
%  
%\begin{equation}  
$ e^{i\alpha}\rightarrow e^{in\alpha}, \;\; 0 \leq \alpha\leq 2\pi$  
%\end{equation}  
%  
and $ n \in \mbox {Z}$  any integer, for the left and right lepton  
fields one takes the transformations (for each generation)  
\begin{equation}  
\psi_L \rightarrow e^{-i\alpha}\psi_L, \qquad  
\psi_R \rightarrow e^{-2i\alpha}\psi_R  
\end{equation}  
(or \mbox {Y} = -1 for $\psi_L$, and \mbox {Y} = -2 for $\psi_R$),  
in order to obtain the correct electric charges of the leptons  
(in units of the elementary charge $e$).  
With the same purpose one sets for the quark fields  
\begin{equation}  
u_L \rightarrow e^{i\alpha\over 3}u_L\, , \qquad d_L \rightarrow  
e^{i\alpha\over 3}d_L\, , \qquad  
u_R \rightarrow e^{4i\alpha\over 3}u_R\, , \qquad d_R \rightarrow  
e^{-2i\alpha\over 3}d_R\, , \qquad etc\;  \cdots \; ,  
\end{equation}  
i.e. \mbox {Y} = 1/3 for $u_L,\; d_L$\ and  \mbox {Y} = 4/3 for $u_R$,  
\mbox {Y} = -2/3 for $d_R$. Obviously, these formulae do not fit with the  
irreducible representations of the group $\mbox {U} (1)$ defined as  
\begin{equation}  
\mbox {U}(1) = \{ e^{i\alpha}\ |\ 0 \leq \alpha\leq\ 2\pi \} \;.  
\end{equation}  

In this paper we propose to use as a gauge group for the WS model  
a three-fold  
covering of $\mbox{U}(2)=(\mbox{U}(1)\times\mbox{SU}(2))/\mbox{Z}_2$  
in order to deal with descent  
representations on the fields. We apply the term metaunitary group for this  
three-fold covering and denote it by $\mbox{MU}(2)$.  
In a recent publication Roepstorff and Vehns \cite{Ro} propose a subgroup $G$  
of $\mbox{SU}(5)$ as a gauge group for the {\em standard model} such that  
$G$ appears  
also as a covering of $\mbox{U}(2)$. Our construction for the  
group  $\mbox{MU}(2)$   is motivated by an argument  
of Guillemin and Sternberg~\cite{Gui}  
for the  definition of a two-fold covering $L$ of the general  
linear group, aimed to  
define a representation of the type $g \rightarrow  
\mbox {det}^{1/2}{g}$ of $L$.  
Noting that the Lie algebras of the groups $\mbox {U} (1)$  
and $\mbox{R}$ coincide,  
we may expect that for the description of fields with electric charge  
proportional to $e/3$, a suitable group may be  
a factor group of $\mbox{R}\times\mbox{SU}(2)$.  
In order to give a more general framework, we present  
in Section II the construction  
of a three-fold covering $\mbox{MU}(n)$ of the group $\mbox{U}(n)$.  
In Section III  
we specialize to the case $n=2$ and consider several representations  
of $\mbox{MU}(2)$ and its Lie algebra  
which will turn useful for the description of leptons  
and quarks in a subsequent paper.  
%In Section III  
%we specialize to the case $n=2$ and consider several representations  
%of $\mbox{MU}(2)$  
%which will turn useful for the description of leptons  
%and quarks in a subsequent paper. The corresponding representations  
%of the Lie algebra of $\mbox{MU}(2)$  
%are given in Section IV.  
%  
  
\section{The Metaunitary Group $\mbox{MU}(n)$}  
  
As mentioned above, we are looking for a gauge group of the electroweak  
interactions  
as a suitable factor group of $\mbox{R}\times\mbox{SU}(2)$.   
Following an argument from \cite{Gui} and in order to provide a more  
general framework, we begin with  
$\mbox{R}\times\mbox{SU}(n)$, $n \geq 2$,  
where the  
group composition law reads  
\begin{equation}  
(u,A).(v,B) = (u + v,AB)\, , \qquad u,v \in \mbox{R}  
\, ,\quad A,B\in \mbox{SU}(n)  
\end{equation}  
and consider the subgroup of $\mbox{R}\times\mbox{SU}(n)$ with elements  
\begin{equation} \label{y} 
 \left\{ \left(k{2\pi \over n},e^{-k{2\pi i\over n}} I \right)  
\ |\ k\in \mbox{Z} \right\} \; ,   
\end{equation}  
which is isomorphic to \mbox{Z}. This subgroup is a normal one and  
through the map $T:\ \mbox{R}\times\mbox{SU}(n) \rightarrow \mbox{U}(n)$, 
defined as   
\begin{equation}  
T\, {(u,A)} = e^{iu}A\; ,  
\end{equation}  
we obtain an isomorphism  
$(\mbox{R}\times\mbox{SU}(n))/\mbox{Z}=\mbox{U}(n) $. Indeed 
 \begin{equation}  
T\, { \left( u + k{2\pi \over n},e^{-k{2\pi i\over n}} A \right)}  
= T\, {(u,A)}\;.  
\end{equation}   
The factor group  $\mbox{MU}(n)=(\mbox{R}\times\mbox{SU}(n))/3\mbox{Z}$, 
consisting of the equivalence classes  
\begin{equation}  
[u,A] = \left\{ \left( u + 3k{2\pi \over n},e^{-3k{2\pi i\over n}} A \right)  
\ |\ k\in \mbox{Z} \right\}\;,  
\end{equation}  
we call {\em metaunitary group}.  
Clearly the map  $T:\ \mbox{MU}(n) \rightarrow \mbox{U}(n)$ defines a  
homomorphism onto $\mbox{U}(n)$. Moreover, $T$ defines a three-fold covering  
of $\mbox{U}(n)$. Indeed, the kernel of $T$ as a map acting  
on $\mbox{MU}(n)$ consists of the elements  
\begin{equation}  
\left[0,I\right]\, , \quad \left[ {2\pi\over n},e^{-{2i\pi \over n}} I\right]  
\;\;\; {\rm and} \;\;\; \left[ {4\pi\over n},  
e^{-{4i\pi \over n}} I \right] \;. \label{yy}  
\end{equation}  
Certainly, the group $\mbox{MU}(n)$ is locally isomorphic to $\mbox{U}(n)$  
and $\mbox{SU}(n)\times\mbox{U}(1)$. The same technique is applicable for 
constructing arbitrary $l-$fold covering of $\mbox{U}(n)$.  
\section{Particular Representations of $\mbox{MU}(2)$ and its Lie Algebra}    
 
\subsection{Representations of $\mbox{MU}(2)$}  
We here specialize to the case $n=2$. Consider    
the map $\mbox{ Det}^{1\over 3}    
:\ \mbox{R}\times \mbox{SU}(2) \rightarrow    
\mbox{U}(1)$ defined by    
\begin{equation}    
\mbox {Det}^{1\over 3}\, {(u,A)} =    
e^{2iu\over 3}, \qquad (u,A) \in \mbox{R}\times \mbox{SU}(2)\;.    
\end{equation}    
Due to the property    
\begin{equation}    
\mbox {Det}^{1\over 3}\, {\left(u+3k\pi ,e^{-3k\pi i}A \right)}    
= \mbox {Det}^{1\over 3}\, {(u,A)}    
\end{equation}    
the map $\mbox{Det}^{1\over 3}$ is well defined on $\mbox{MU}(2)$ and gives a    
homomorphism of $\mbox{MU}(2)$ onto $\mbox{U}(1)$.    
For every integer $k$ the mapping    
$\mbox{Det}^{k\over 3}:\ \mbox{MU}(2) \rightarrow \mbox{U}(1)$ given by    
\begin{equation}    
\mbox {Det}^{k\over 3}\, {[u,A]} =    
{\left( \mbox {Det}^{{1\over 3}}\, {[u,A]} \right)}^k\; ,    
\end{equation}    
is also a homomorphism of $\mbox{MU}(2)$ on $\mbox{U}(1)$. For {k = 3}    
\begin{equation}    
\mbox{Det}{[u,A]} = \mbox{det}\circ T\, {[u,A]}    
\end{equation}    
where ``det" stands for the usual determinant.    
Using the maps {T} and $\mbox{Det}^{k\over 3}$ we define the homomorphisms    
$T^k :\ \mbox{MU}(2) \rightarrow \mbox{U}(2)$ by    
\begin{equation}    
T^{k}\, {[u,A]} = \mbox{Det}^{k\over 3}\, {[u,A]} \, T\, {[u,A]} =    
e^{iu \left(1+{2k\over 3} \right) }A\;\;.    
\end{equation}    
Some particular cases are    
\begin{eqnarray}    
&&T^0\, {[u,a]} = T\, {[u,A]} = e^{iu}A\;, \qquad    
\ \ \ T^{-2}\, {[u,A]} =  e^{-{iu\over 3}} A\;, \label{eq:17}    
\end{eqnarray}    
\begin{eqnarray}    
&&\mbox{ Det}^{1\over 3}\, {[u,A]}= e^{ 2iu \over 3}\, ,    
\qquad \mbox{ Det}^{-{2\over 3}}\, {[u,A]} = e^{ -{4iu \over 3}}\, ,    
\qquad \mbox{ Det}\, {[u,A]} = e^{2iu}\;. \label{eq:18}    
\end{eqnarray}    

The one-parameter subgroup of $\mbox{MU}(2)$,    
\begin{eqnarray}    
\mbox{MU}_{\mbox{em}}(1) = \left\{    
\left[-{\alpha \over 2}, \left( \begin{array}{cc}    
e^{i\alpha \over 2}     &0      \\    
0       &e^{i\alpha \over 2}    \end{array} \right) \right]    
\in \mbox{MU}(2) \ |\  \alpha \in \mbox{R}    
\  \right\}\;,    
\end{eqnarray}    
has the meaning of the group generated by the electric charge generator    
\begin{equation}    
\mbox{Q} = \frac{1}{2}\mbox{Y} + \mbox{I}_3 =  \frac{1}{2} I +     
\frac{1}{2} \sigma_3 \;,   
\end{equation}    
where    
$$I =  \left( \begin{array}{cc}    
1 &0  \\    
0 & 1 \end{array}    
\right)  \quad {\rm and} \quad   
 \sigma_3 =  \left( \begin{array}{cc}    
1 &0  \\    
0 &-1 \end{array}    
\right) \;.$$    
The image of $\mbox{Q}$ in each representation of $\mbox{MU}_{\mbox{em}}(1)$    
in a space $\mbox{V}$ is identified with the electric charge in this    
representation, the eigenvalues $q_i$ of $\mbox{Q}$ being identified with    
the charges of the corresponding eigenvectors from $\mbox {V}$.    

Let $e_1,e_2$  be a basis in ${\Bbb C}^2$. The second exterior degree    
$\Lambda^2 {\Bbb C}^2$ is a one-dimensional complex space generated by    
$e_1 \wedge e_2$ which carries the representations    
$\mbox{Det}^{k\over 3}$ for different $k$.    
Let $a = {u\,e_1 + v\,e_2}\in {\Bbb C}^2$ and     
$w\,e_1\wedge e_2 \in \Lambda ^2 {\Bbb C}^2$.    
Using the notation ${\cal M}(\alpha ),\; 0 \leq \alpha \leq 2\pi$, for  
$\mbox{MU}_{\mbox{em}}(1)$, one finds in the representations    
(\ref{eq:17},\ref{eq:18})    
\begin{eqnarray}    
1) \ \ &&  T\,\left[{{\cal M}(\alpha)}\right]    
%{ \left[-{\alpha \over 2}, \left( \begin{array}{cc}    
%\scriptsize    
%e^{i\alpha \over 2} &\scriptsize 0  \\    
%\scriptsize 0 & \scriptsize e^{i\alpha \over 2}  \end{array}    
%\right) \right] }    
= \left( \begin{array}{cc}    
1       &0      \\    
0 &e^{-i\alpha }  \end{array}    
 \right)\;,    \qquad    
\ \ \ \ \ \; Q_{T} =  \left( \begin{array}{cc}    
0 &0  \\    
0 &-1 \end{array}    
\right)\;,  \ \ \ \qquad q_{u} = 0\;,\ q_v = -1\; .  \label{frep} \\[1.5ex]    
2) \ \ &&  T^{-2}\,\left[{{\cal M}(\alpha)}\right]    
% \left[-{\alpha \over 2}, \left( \begin{array}{cc}    
%\scriptsize e^{i\alpha \over 2} &\scriptsize 0  \\    
%\scriptsize 0 &\scriptsize e^{i\alpha \over 2}  \end{array}    
%\right) \right] }    
= \left( \begin{array}{cc}    
e^{2i\alpha \over 3}    &0      \\    
0       &e^{-i\alpha \over 3 }  \end{array}    
\right)\;,   \quad \,   
\ Q_{T^{-2}} =  \left( \begin{array}{cc}    
{2\over 3}      &0      \\    
0       &-{1\over 3}    \end{array}    
\right)\;, \qquad q_{u} = {2\over 3},\ q_v = -{1\over 3}\; .  \\[1.5ex]    
3) \ \ &&  \mbox{Det}^{{1\over 3}}\,\left[{{\cal M}(\alpha)}\right]    
%\left[-{\alpha \over 2}, \left( \begin{array}{cc}    
%\scriptsize e^{i\alpha \over 2} &\scriptsize 0  \\    
%\scriptsize 0 &\scriptsize e^{i\alpha \over 2}  \end{array}    
%\right) \right] }    
= e^{-{i\alpha \over 3}}\;, \qquad  \qquad    
\ \ \ \ \ Q_{\mbox{Det}^{{1\over 3}}} = -{1\over 3}\;,    
\qquad  \qquad \ \ \, q_{w} =- {1\over 3} \;.  \\[1.5ex]    
4) \ \ &&  \mbox{Det}^{-{2\over 3}}\,\left[{{\cal M}(\alpha)}\right]    
%\left[-{\alpha \over 2}, \left( \begin{array}{cc}    
%\scriptsize e^{i\alpha \over 2} &\scriptsize 0  \\    
%\scriptsize 0 &\scriptsize e^{i\alpha \over 2}  \end{array}    
%\right) \right] }    
= e^{2i\alpha \over 3}\;, \qquad  \qquad \ \    
 \ \ Q_{\mbox{Det}^{-{2\over 3}}} = {2\over 3}\;,   
 \qquad \;  \qquad \ \ \, q_{w} = {2\over 3} \;.  \\[1.5ex]    
5) \ \ &&  \mbox{Det}\,\left[{{\cal M}(\alpha)}\right]    
%\left[-{\alpha \over 2}, \left( \begin{array}{cc}    
%\scriptsize e^{i\alpha \over 2} &\scriptsize 0  \\    
%\scriptsize 0 &\scriptsize e^{i\alpha \over 2}  \end{array}    
%\right) \right] }    
= e^{-i\alpha }\;,  \qquad \qquad \ \    
 \ \ \ \ \ \; Q_{\mbox{Det}} = -1\;, \ \ \;  \qquad \ \ \,    
\qquad q_{w} =- 1 \;. \label{lrep}    
\end{eqnarray}    
The choice of these representations is justified by the reduction of    
$\mbox{MU}(2)$ to $\mbox{MU}_{\mbox{em}}(1)$ in analogy with the reduction of    
$\mbox{SU}(2)\times\mbox{U}(1)$ to $\mbox{U}_{\mbox{em}}(1)$ in the WS model.    
Then a direct sum of one-dimensional representations of    
$\mbox{MU}_{\mbox{em}}(1)$ appears, each of them with a fixed electric charge.    

\subsection{Representations of the Lie Algebra of $\mbox{MU}(2)$}    
    
The groups $\mbox{MU}(2)$ and    
$\mbox{R} \times \mbox{SU}(2)$ are locally isomorphic and    
one has for their Lie algebras    
\begin{equation}    
Lie\, \mbox{MU}(2) = \mbox{R} \oplus Lie\, \mbox{SU}(2) \;.    
\end{equation}    
Accordingly, a set of four generators for $\mbox{MU}(2)$    
is given by    
\begin{equation}    
X^a = \left( 0,{\sigma ^a \over 2} \right)\, , \ a=1,2,3 \;,     
{\rm and}  \qquad X = \left( -{1\over 2},0 \right)\; ,    
\end{equation}    
where $\sigma^a$ are the Pauli matrices and    
$$ [X^a,X^b] = i\varepsilon ^{abc} X^c, \qquad [X^a,X] = 0 \;.$$    
The subgroups of $\mbox{MU}(2)$, generated    
by $X^a$ and $X$, are    
% `   
\begin{eqnarray}    
 G_{X^a}(t)& = &\left\{ \left[0,e^{{i\sigma ^a \over 2}t} \right] \    
|\ t \in \mbox{R} \right\}, \ a=1,2,3\, , \\[1ex]    
 G_{X}(t)& = &\left\{ \left[-{1\over 2}t,I \right] \    
|\  t \in \mbox{R} \right\}\;.    
\end{eqnarray}    
Each representation $T$ of the group $\mbox{MU}(2)$ generates a representation    
$T_*$ of its Lie algebra. For the particular representations     
(\ref{frep})-(\ref{lrep})  defined    
in the previous    
subsection one finds for the generators $X^a$ and $X$    
\begin{eqnarray}   
1) \  &&T\, {\left[0,e^{{i\sigma ^a \over 2}t}  \right]} =   
e^{{i\sigma ^a \over 2}t}\, , \qquad T\, {\left[-{1\over 2}t,I \right]} =   
e^{-{i\over 2}t}I \;. \\[1ex] \label{eq:200}   
&&\ \ \ T_* (X^a) = -i{d\over dt} e^{{i\sigma ^a \over 2}t} \ _{|t=0}  =   
{\sigma ^a \over 2}\;. \\[1ex]   
 &&\ \ \ T_* (X) = -i{d\over dt} e^{-{i \over 2}t}I\ _{|t=0}   =   
-{I \over 2} \;. \\[1ex] && \nonumber  \\    
2)\  &&T^{-2}\, {\left[0,e^{{i\sigma ^a \over 2}t} \right]} =   
e^{{i\sigma ^a \over 2}t}\, , \qquad T^{-2}\, {\left[-{1\over 2}t,I \right]} =   
e^{{i\over 6}t}I\;.  \\[1ex]   
 &&\ \ \ T^{-2}_* (X^a) = -i{d\over dt} e^{{i\sigma ^a \over 2}t} \ _{|t=0}  =   
{\sigma ^a \over 2} \;.\\[1ex]   
 &&\ \ \ T^{-2}_* (X) = -i{d\over dt} e^{-{i \over 2}t}I \ _{|t=0}  =   
{I \over 6} \; . \\[1ex] && \nonumber\\    
3)\  &&\mbox{Det}^{{1\over 3}}\, {\left[0,e^{{i\sigma ^a \over 2}t}   
\right]} = 1 \, , \qquad \mbox{Det}^{{1\over 3}}\, {\left[-{1\over 2}t,I   
\right]} = e^{-{i\over 3}t}\;.  \\[1ex]    
&&\ \ \ \mbox{Det}^{{1\over 3}}_* (X^a) =   
 -i{d\over dt} 1 \ _{|t=0} = 0 \;. \\[1ex]    
&&\ \ \ \mbox{Det}^{{1\over 3}}_* (X) = -i{d\over dt}    
e^{-{i \over 3}t}  \ _{|t=0} = -{1 \over 3} \;. \\[1ex] && \nonumber \\   
4)\  &&\mbox{Det}^{-{2\over 3}}\, {\left[0,e^{{i\sigma ^a \over   
 2}t} \right]} = 1\, , \qquad \mbox{ Det}^{-{2\over 3}}\, {\left[-{1\over   
2}t,I \right]} = e^{{2i\over 3}t} \;. \\[1ex]    
&&\ \ \ \mbox {Det}^{-{2\over 3}}_*   
 (X^a) = -i{d\over dt} 1 \ _{|t=0}  = 0\;.  \\[1ex]    
&&\ \ \ \mbox {Det}^{-{2\over 3}}_* (X)   
 = -i{d\over dt} e^{{2i \over 3}t}  \ _{|t=0} = {2 \over 3}\;. \\[1ex] &&     
\nonumber \\   
5)\ &&\mbox {Det}\, {\left[0,e^{{i\sigma ^a \over 2}t} \right]} = 1 \, , \qquad   
\mbox{ Det}\, {\left[-{1\over 2}t,I \right]} = e^{-it}\;. \\[1ex]    
&&\ \ \ \mbox   
 {Det}_* (X^a) = -i{d\over dt} 1 \ _{|t=0}  = 0  \;.\\[1ex]    
&&\ \ \ \mbox {Det}_* (X) =   
 -i{d\over dt} e^{-it} \ _{|t=0}  = -1 \;. \label{eq:300}    
\end{eqnarray}    
These representations of $Lie\, \mbox{MU}(2)$ will be used in a    
subsequent paper for the explicit form of the covariant derivatives    
of the fields in a model based on the gauge group $\mbox{MU}(2)$.    
    
\newpage  
\vskip 1cm    
\begin{center}    
{\bf ACKNOWLEDGMENTS}    
\end{center}    
\vskip 0.25cm    
One of us (V.R.) is grateful to Prof. R. Kerner and the    
Laboratoire de Physique des Particules, Universit\'e Pierre et    
Marie Curie, Paris, for their hospitality during a visit when     
this work was started.  This work was partly supported (I.V.) by the   
DOE Research Grant under Contract No.  
De-FG-02-93ER-40764.

\end{document}